# Real time, cross platform visualizations with zero dependencies for the N-body package REBOUND

AUTHOR

Hanno Rein ✉ ⓘ

AFFILIATIONS

Department of Physical and Environmental Sciences, University of Toronto

Department of Astronomy and Astrophysics, University of Toronto

Department of Computer Science, University of Toronto

**Abstract**

Background

Visualizations have become an indispensable part of the scientific process. A vibrant ecosystem of visualization tools exists, catering to a wide variety of different needs. Real-time visualizations of numerical simulations offer scientists immediate feedback about the status of their simulations and can also be valuable educational and public outreach tools.

Current Challenges

Developing a visualization tool with support for different operating systems, CPU/GPU architectures, and programming languages can be a challenge. It is common to use one or more UI toolkits or libraries to act as abstraction layers and hide the underlying complexity. Whereas external libraries greatly simplify the initial programming effort, we argue that relying on them introduces new dependencies and problems, such as a higher barriers to entry for new developers and users, and uncertainty regarding long-term support.

Proposed Solution

In this paper we present a new approach for real-time visualizations which we have implemented for the N-body package REBOUND (Rein and Liu 2012). We propose to use a web browser to handle GPU accelerated rendering. This enables us to offer 3D, interactive visualizations of simulations running natively on all major operating systems. What makes our new approach unique is that we achieve this without the need for any external libraries. We utilize WebAssembly and emscripten to reuse existing OpenGL visualization code. Using communication via HTTP and a custom built-in web server, we are able to provide both local and remote real-time visualizations. In addition to the browser based real-time visualization, our approach offers other additional operating modes, including simulations running entirely within the browser, visualizations within jupyter notebooks, and traditional standalone visualizations using OpenGL. We focus on the implementation in REBOUND but the concepts and ideas discussed can be applied to many other areas in need of scientific and non-scientific real-time visualizations.

Materials, License, Conflicts

# 1 Introduction

Visualizations are a crucial part of the scientific process. There are many popular tools to make two or three dimensional graphs such as gnuplot (Williams and Kelley 2013), matplotlib (Hunter 2007), or yt (Turk et al. 2011) for rendering volumetric and particle data. Many specialized tools exist as well, for example Williams et al. (2022) describe a novel approach for an interactive user interface for the WorldWide Telescope. The above examples demonstrate the diversity of visualization needs and the tools that have been developed to serve those needs. However, it would be impossible to review all the software that is available.

Developing software that involves any sort of graphical interface can be a challenge. This is especially the case if the goal is to provide a tool that works on different operating systems and supports a variety of graphics hardware. There are two ways to approach the problem of cross-platform graphics that are commonly employed:

- One can write tailored graphic routines for each platform. This requires significant resources, both for development and maintenance. Whereas this approach might be an option for large projects (think of a game developed by a large studio), it is typically not within the realm of possibilities for small scientific software packages. If resources are finite, one might end up with limited support of only a few platforms, or with some platforms enjoying more features than others. Mobile applications are an example where this approach is often used and as a result iOS and Android versions of the same application do not necessarily share the same features.
- Alternatively, one can rely on a cross-platform libraries, frameworks, or APIs such as QT, Unity, OpenGL, or Vulkan. These make cross-platform development much more straightforward but at the cost of adding a dependency. Scientific software, in particular if it involves simulating scientific processes, is mostly distributed as source code because researchers want the ability to study and modify it. Compiling source code that depends on external libraries into an executable can be a major obstacle for new developers and users alike. Furthermore, switching from one library to another is not trivial. This can become a problem when external libraries don't get updated or become deprecated. As an example, many games and visualization tools depend on OpenGL which is now considered deprecated on MacOS. Furthermore, for commercial graphic toolkits such as Unity, license agreements might change at any time. In short, the development is now highly dependent on whichever external component was chosen for the project.

This paper presents an alternative approach which we think has several important advantages over the two approaches above, especially for small scientific projects:

- Dependency free. There are no external libraries required.[1]
- Cross platform. Any platform that has a reasonably modern browser can be used to render visualizations. This includes both desktop and mobile operating systems: Linux, MacOS, Windows, iOS, Android.
- Remote visualizations. Using port-forwarding via SSH, one can run a simulation on one computer, say a node of a computing cluster, and visualize it in real-time on another workstation.
- Future-proof. Because the approach relies on open web technologies that are supported by all major browsers, we find it likely that this approach will continue to work for many years to come without requiring much maintenance in the same way one can still view websites that were developed decades ago.

In the following sections, we will describe how this approach works in detail.

## 2 Operating modes

A key feature of our approach is its flexibility. Figure 1 shows the three different operating modes that are possible. Modules with the same colour in the figure make use of mostly the same source code. The high level of code-reuse is possible because C/C++ code can be compiled with emscripten to WebAssembly which can then be interpreted by a web browser at almost native speed. In summary, the different modes work as follows:

1. In the **standalone mode**, the visualization is making use of OpenGL. This traditional approach has been used by both visualization tools and games for decades. It requires no browser, but in addition to a C compiler, OpenGL APIs need to be available, and GLFW libraries need to be installed. This mode provides the best performance.
2. In the **hybrid mode**, when a REBOUND simulation is started, a web server is automatically started on a separate thread. The web server serves both visualization code and simulation data to a web browser. The visualization is then done in the browser. The server and browser do not need to run on the same machine. This mode does not require any external libraries to be installed for compilation. It provides a low entry barrier for developers and users (*it just works*). This mode constitutes the novel visualization concept that we describe in this paper.
3. In the **browser mode**, not only the visualization is handled by the browser but also the simulation itself. This mode requires no external libraries nor any server. The emscripten compiler is used to bundle everything that is needed by the browser (HTML, CSS, JavaScript, WebAssembly, image data) into one single HTML file that can then be served as a static website or directly be opened locally with a web browser.

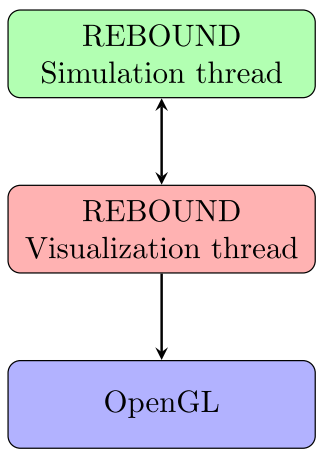
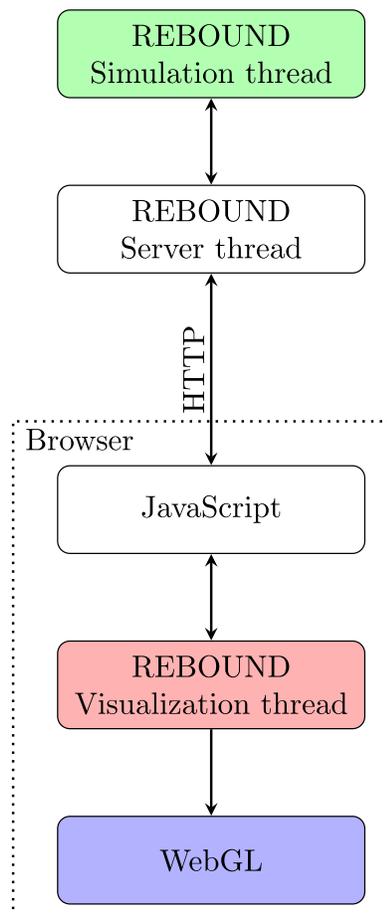
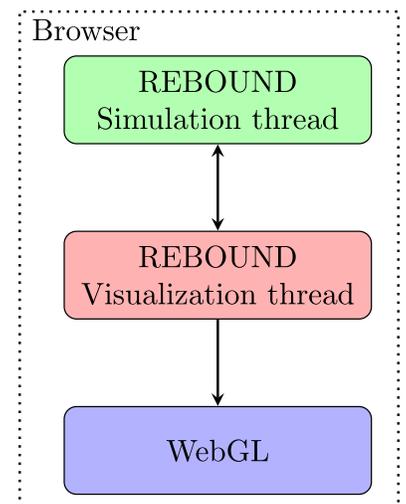

Figure 1: An overview of the three different operating modes. In the standalone mode (left), the visualization is provided by OpenGL. In the hybrid mode (middle), the browser streams data from a web server which is then visualized using WebGL. In the browser mode (right), the simulation and the visualization are handled solely within the browser. Modules with the same colour (green, red, blue) make use of the same source code.

Both the standalone and the browser mode have been employed widely by other tools. The innovation of this paper is the hybrid mode which in many ways represents the best of both worlds. In the case of REBOUND, a user simply downloads the source code and compiles it using any C compiler[2]. Because there are no dependencies on external libraries, this does not require any configure scripts, installing system-wise third party packages, setting up environment variables, or other complicated steps. The user can setup or run a simulation without even considering whether a visualization might be useful. If the user at some point decides that it might be useful, then they can simply open a web browser and point it towards the simulation's dormant web server to start visualizing the simulation and get immediate visual feedback regarding the simulation.

Because we already have the visualization code running in a web browser, the additional task of running entire simulations in the web browser is straightforward. Figure 2 shows an example of REBOUND's browser mode. After clicking on the figure (this only works if you are reading the HTML version of this article), an N-body simulation of a self-gravitating disk can be seen. The simulation is running in real-time in the browser. The rendering is using WebGL. The console output is also shown. The visualization is interactive: drag to rotate, shift+drag or scroll to zoom, press the space bar to pause. The figure is included in this document using an `<iframe>` HTML tag. A single file includes the bundled up HTML, CSS, JavaScript, WebAssembly as well as some small image assets and is less than 500 kB in size. For comparison, a simple screenshot of the simulation in PNG format would be 200 kB in size[3]. The REBOUND documentation makes extensive use of the browser mode, offering users the ability to run all examples directly in the browser and thus showcasing the ability of the software package. Note that the exact same simulation can be compiled with a normal C compiler and run in the stand-alone mode showing the same visualization but using OpenGL instead of WebGL.

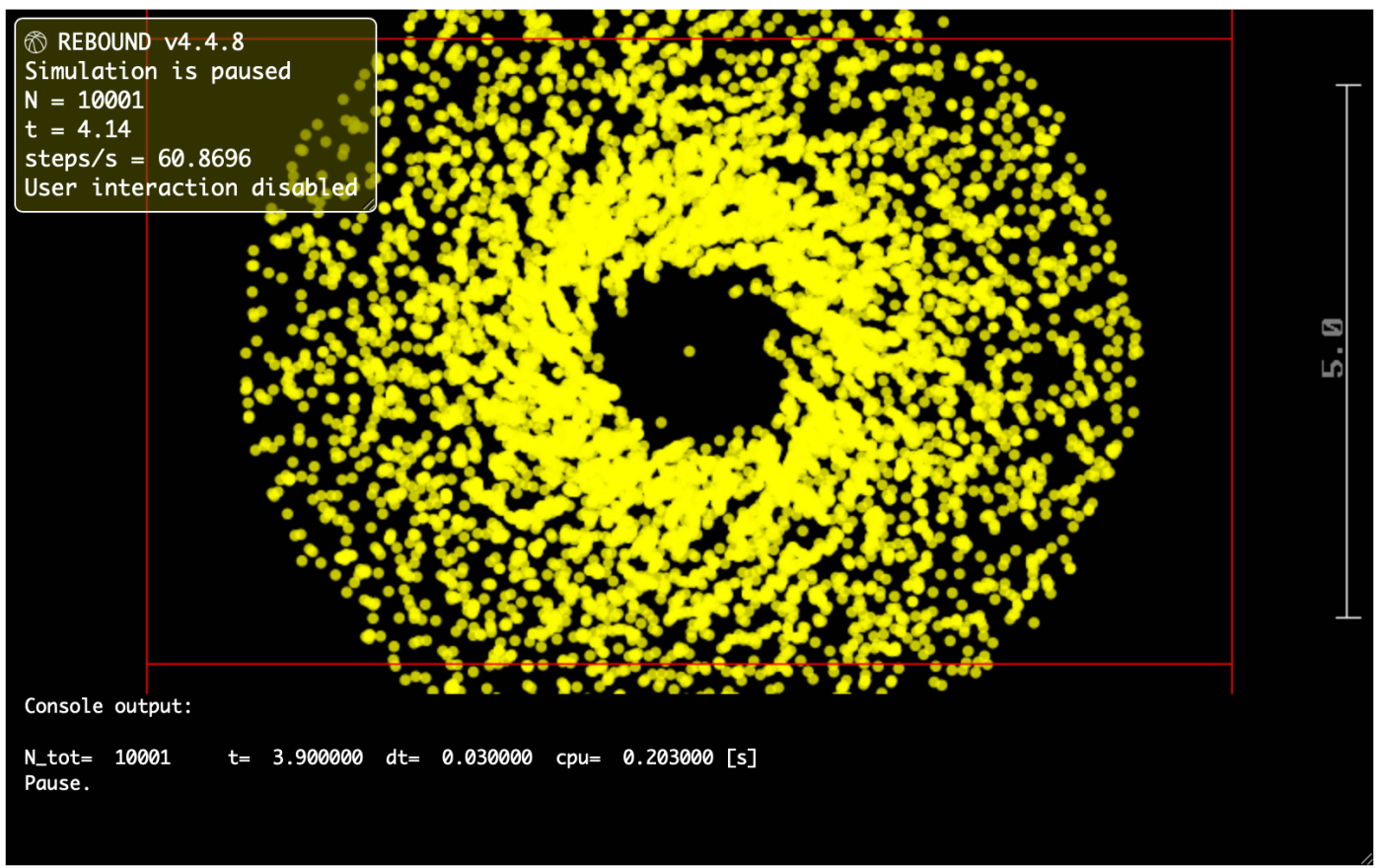

Figure 2: This figure shows a screenshot of an interactive, real-time visualization of a self-gravitating disk where the equations of motion are being integrated with the N-body code REBOUND. To achieve this, REBOUND has been compiled with emscripten

to WebAssembly allowing it run in the browser. This is what we refer to as the *browser mode*. For the interactive version of this figure, see the HTML version of this paper.

## 3 Web Browser, WebGL, WebAssembly

We rely on a web browser for all graphic related functionality in the hybrid and browser modes. There are different ways to render visualizations within a browser. For example, one can use HTML, CSS and JavaScript. A popular JavaScript visualization library is D3.js.

In recent years, browsers have also gained features that allow them to efficiently render high quality graphics and make use of GPU acceleration. Although the performance is not (yet) on par with native desktop implementations, it comes very close and is sufficient for most applications (see also our performance tests below). We make use of these features and write GPU accelerated visualizations in the browser using WebGL, a JavaScript API that provides similar functionality to OpenGL. All major browsers support WebGL version 2.

We also make use WebAssembly, a binary instruction format that can be executed by a browser's virtual machine at almost native speeds. Using the emscripten compiler toolchain, one can compile existing C or C++ code to WebAssembly. Emscripten also converts OpenGL to WebGL so that we don't need to explicitly write WebGL code. With minor adjustments, many games and visualization tools that use OpenGL on a desktop can thus also be used in a browser. Similar to JavaScript libraries, programs in WebAssembly format are simply static files and don't require the user to install anything, avoiding any of the issues that come with external libraries. There are several advantages of focusing on WebAssembly and emscripten for the in-browser rendering code.

First, we can reuse our existing C code that already uses OpenGL for rendering, with only minimal changes (see Section 7). This makes it easy for existing projects to migrate to browser-based rendering. Other programming languages such as python are also beginning to be supported in browsers (Droettboom et al. 2021).

Second, not only can we reuse graphic related code, but also any other C code. In the case of REBOUND, the code running in the browser uses the same input and output routines as the N-body simulation itself. By reusing these routines, it becomes very easy to unpack visualization data from a file or data stream, thus allowing us to send back and forth data over the network - which is what we need for the hybrid mode. Specifically, we are using REBOUND's Simulationarchive binary format (Rein and Tamayo 2017) to send data to the browser. Although the data transfer using the Simulationarchive format is easy to setup, it is not optimal as the Simulationarchive is not compressed and the full simulation is sent every frame, rather than only the changes.

Third, we can even run an entire simulation in a web browser (we call this browser mode), generating data that is being visualized on-the-fly. This allows visualizations to be included in a static website such as Figure 2 in this paper.

Fourth, we are not restricting ourselves to the browser ecosystem. We can also run visualizations locally, without the need for emscripten, WebGL, or a browser, by simply calling OpenGL APIs the old fashioned way. This makes development easier and allows advanced users who are comfortable installing libraries to take full advantages of the efficiencies and all the features that a native application provides.

There are also some drawbacks to using WebGL and WebAssembly. First, a WebAssembly compiler needs to be installed when a user wants to modify the visualization code. WebAssembly compilers such as emscripten are still relatively new and are not yet as widely available and as well tested as many C compilers. Second, debugging WebAssembly code might pose a challenge given that the compiled code makes it harder to inspect the code in the browser due to a lack of good debugging tools. And third, using WebAssembly can impact accessibility. For example text rendered via WebGL might not be picked up by the operating system or browser accessibility tools such as screen readers (the same is true for OpenGL). For this reason, we do not render text via WebGL in our implementation and instead use HTML for this.

## 4 Performance

To measure and compare the performance (refresh rate) of the three different operating modes, we run the self-gravitating disk simulation from Figure 2 with $N=10^4$ as well as $N=10^5$ particles and measure the average frames per second we achieve with each visualization mode. The tests were run on a 2020 MacBook Air (M1). Safari was used as a browser but the measurements were almost identical for Chrome and Firefox. The results are shown in Table 1.

For small simulations ($N=10^4$), the performance is comparable for all three modes (42 - 71fps). The hybrid mode is the slowest, which can be explained by the additional data transfer over http. For larger particle numbers ($N=10^5$) the native OpenGL visualization performs significantly better (41fps) than the hybrid or browser mode (2.2fps and 3.7fps). The hybrid mode is slow because at every frame 12.8MB of data need to be transferred over http. The browser mode is slow for a different reason: the simulation and the visualization run on the same thread. Thus, while the integrator is computing the next timestep, the visualization appears frozen. This can by verified by pausing the simulation, in which case the visualization performs significantly better (29fps). WebAssembly and emscripten support multithreading and shared memory access across threads via `SharedArrayBuffer`s. However, this feature is currently not used in the browser mode because it is gated behind a Cross Origin Opener Policy[4] which makes deployment more difficult.

In summary, for small particle numbers ($N \lesssim 10^4$), there is no significant performance difference. For large simulations with millions of particles, the present approach is not feasible due to its single threaded implementation. However, note that the main use case of REBOUND is running small N-body simulations of planetary systems with few ($N < 10$) particles.

Also shown in Table 1 is the runtime required to integrate for 10 code units. The fastest simulations are of course those without any visualization. The runtime increases only very slightly (10%) when using the hybrid mode. Simulations using the OpenGL mode or the browser mode have more overhead. This results in smoother visualizations (higher fps) at the cost of somewhat longer runtimes.

Table 1: Performance of the different visualization modes using the self-gravitating disk simulation from Figure 2 as an example.

| Mode | $N=10^4$ **particles** | | $N=10^5$ **particles** | |
| --- | --- | --- | --- | --- |
| | refresh rate | runtime | refresh rate | runtime |
| OpenGL | 71 fps | 9.1s | 41 fps | 1m 50s |
| Hybrid | 42 fps | 4.7s | 2.2 fps | 1m 22s |

| Mode | $N = 10^4$ **particles** | | $N = 10^5$ **particles** | |
|---|---|---|---|---|
| | refresh rate | runtime | refresh rate | runtime |
| Browser | 54 fps | 7.3s | 3.7 fps | 1m 38s |
| No visualization | N/A | 4.4s | N/A | 1m 17s |

# 5 Web Server

As we delegate rendering functionality to the browser, we somehow need to get the data that we wish to visualize into the browser. For the browser mode, one can simply use static files which the browser can access, or even include the data within a single HTML file (this is what has been done to create Figure 2).

For the hybrid mode, which visualizes a remote simulations in real-time, we need to send the data to the browser when it is needed or when it is available. Given that the rendering takes place in a browser, we use HTTP requests and a web server for this data transfer.

There are many libraries and packages that implement a web server. Examples are the Apache HTTP Server Project, or python's default `http.server` module. However, one of our initial goals was to avoid any external libraries. So instead of using an existing http library, we implement our own. This may seem like a daunting task, but the reference implementation we provide in REBOUND is only about 500 lines of C code and works natively on all major operating systems including Linux, MacOS, and Windows[5]. This is possible because our web server doesn't need to support security features - it is not intended for use on public networks. It just performs the task of getting data to our rendering code in the browser. Since we are implementing both the front and backend, we have tight control over what kind of requests the server needs to be able to respond to and we can simply ignore the rest.

In addition to sending data from the simulation to the browser, the web server can also receive commands and data from the browser. For example, in REBOUND these commands allow the user to pause or quit a running simulation, or step through simulations manually. We furthermore make use of HTTP POST requests to send screen captures from the browser to the server. The user may opt to save these screenshots on the server side, for example to later combine individual frames into an animation. To further facilitate the rendering of complex animations, we also implement a way for the server side to push the view matrix and other visualization setting to the browser and override any client side user interaction. The beginning of the simulation in Figure 2 demonstrates this: the camera automatically zooms in and rotates by 90 degrees at the beginning of the simulation before yielding control over the visualization to the user.

# 6 Jupyter Notebooks

Although REBOUND is written in C, it also comes with a python interface, allowing users to more easily setup and run N-body simulations that do not require modifying and compiling the REBOUND source code itself. Jupyter notebooks (Kluyver et al. 2016) are a popular web-based interactive environment to edit and run python code. Given that we now have a web-based visualization tool for REBOUND, we can easily incorporate it into a Jupyter notebook using the hybrid mode. A user can start an interactive visualization of a simulation within a Jupyter notebook by simply calling the `.widget()` function on any simulation object. All this function does is insert an iframe in the notebook and point it to the simulation's web server using the `Iframe` class of `IPython.display`. Because this is using the hybrid mode, this workflow does also not rely on any external libraries, neither on the C nor the python/IPython side.

# 7 Step by step guide

Here, we summarize the modifications we had to make to the existing OpenGL visualizations in REBOUND in order to add the new hybrid and browser modes. The goal of this section is to help other developers who want to adopt a similar visualization strategy. We assume that an existing desktop visualization based on OpenGL and written in C already exists.

- The OpenGL shaders in the browser use an Embedded Systems (ES) version. To continue supporting both the desktop and the ES version, we hard code this difference with a pre-compiler directive.
- Because the capabilities of the ES version are more restricted, some features have been disabled with a pre-compiler directive. For example, we disable multi-sampling.
- We disable the OpenGL based rendering of on-screen text. Rendering text is non-trivial in pure OpenGL[6]. However, it is very easy in a browser with the additional benefit of better supporting accessibility tools such as screen readers. Rendering text involves updating an HTML element by calling javascript code from C with emscripten's `EM_JS` API.
- We also need to adjust the rendering loop. On the desktop version the rendering loop is using `glfwGetTime()` to call the rendering function at a fixed maximum frame rate. On the emscripten side, we use the `emscripten_request_animation_frame_loop()` API call instead.

After making the above changes, one should be able to compile the code with emscripten and run the simulation together with the visualization in the browser (i.e. what we call the browser mode). This is a good opportunity to fix any new bugs or rendering issues that arose from the above changes.

- To get the hybrid mode working, we need to implement a web server. The reference implementation in REBOUND can be used as a starting point. The only two required capabilities of the web server are to serve a static HTML file (which contains the WebAssembly code for running the browser based visualization) and send updated simulation data on request.
- The web server is started on a separate thread when a simulation is created. It initially checks if it can find the static html file the browser needs and, as a fallback, downloads it from the internet if does not exist. The latter is not strictly necessary but allows users to avoid either compiling or downloading the file manually.
- What simulation data to send to the browser at every frame depends on the nature of the simulation. In REBOUND, we use the Simulationarchive to serialize and send the entire simulation. This way, we don't need to implement complicated logic for changing simulation states as the Simulationarchive contains all information to reconstruct the simulation. Alternatively, we could only send the updated particle data every timestep, but then we would need to handle special cases, for example when the particle number changes which involves reallocating various arrays both on the CPU and GPU side. For other simulations, e.g. a hydrodynamic simulation with a fixed size grid, the latter option would achieve better performance.
- While the simulation is getting updated during a timestep, it locks a Mutex such that the server thread doesn't send a partially updated simulation. That could lead to race conditions and memory corruption.

- On the browser side, we create a new simulation and visualize it the same way as we do for the browser mode. However, instead of initializing the simulation with particle data directly in the browser, we request a Simulationarchive from the server, then use that to update the in-browser simulation. Once done, we pause for a few milliseconds before requesting another Simulationarchive. The visualization gets updates in the next rendering call.

At this point, all the components in order for the hybrid mode to work are in place. The next steps are to polish the HTML file that emscripten uses as a template so that it can show some status information, render any onscreen text, and if desired send user interactions back to the server. There are also various failure modes that need to be handled such as a failed attempt to connect the server[7].

# 8 Conclusions

In this paper, we have presented a flexible framework to add real-time visualization to a scientific simulation package. Our approach provides GPU accelerated visualizations on all major operating systems with zero dependencies on external libraries. We achieve this by relying on a web browser for rendering, and implementing a web server from scratch. Our approach is compatible with modern web-based environments such as Jupyter notebooks and also allows interactive visualizations to be embedded in static websites such as scientific publications, documentation, and blogs.

A reference implementation of our approach is provided in the REBOUND N-body package (Rein and Liu 2012). The most recent development version is available on GitHub. The version of REBOUND that was used for this paper (4.4.10) is also available as an archive on Zenodo. We were able to reuse the majority of the rendering code from the existing OpenGL based visualization in REBOUND by converting it to WebGL and WebAssembly with emscripten. We believe our approach is general enough that it will be applicable to a wide variety of other software packages which currently lack the possibility of real-time visualizations and we provide a step by step guide to get started.

There are many possibilities for further improvements. We would like to highlight one which would make this framework even more accessible. Pyodide (Droettboom et al. 2021) is a port of CPython to WebAssembly which makes it possible to install and run Python packages in the browser. In contrast to the browser mode discussed above, with Pyodide a user can interpret python code interactively directly in the browser. REBOUND already is available as a Pyodide package and one can thus code up and run a custom REBOUND simulation entirely from within the browser. However, the visualization framework presented in this paper still has to be ported. The main difficulty is that communication between the visualization thread and the simulation thread needs to be changed (we can no longer communicate over network ports) and the support of threads is somewhat restricted in WebAssembly (UI changes need to come from one thread and thus needs to be shared with the python console).

# Authorship

**Hanno Rein**: Conceptualization, Implementation, Writing.

# License

## Conflict of Interest

The author declares that there are no other competing interests.

## References

Droettboom, Michael, Roman Yurchak, Hood Chatham, Dexter Chua, Marc Abramowitz, casatir, Jason Stafford, et al. 2021. "Pyodide/Pyodide:" Zenodo. https://doi.org/10.5281/zenodo.5156931.

Hunter, J. D. 2007. "Matplotlib: A 2D Graphics Environment." *Computing in Science & Engineering* 9 (3): 90–95. https://doi.org/10.1109/MCSE.2007.55.

Kluyver, Thomas, Benjamin Ragan-Kelley, Fernando Pérez, Brian Granger, Matthias Bussonnier, Jonathan Frederic, Kyle Kelley, et al. 2016. "Jupyter Notebooks—a Publishing Format for Reproducible Computational Workflows." *Positioning and Power in Academic Publishing: Players, Agents and Agendas*, 87.

Rein, H., and S. -F. Liu. 2012. "REBOUND: an open-source multi-purpose N-body code for collisional dynamics" 537 (January): A128. https://doi.org/10.1051/0004-6361/201118085.

Rein, H., and D. Tamayo. 2017. "A New Paradigm for Reproducing and Analysing n-Body Simulations of Planetary Systems." *MNRAS* 467 (January): 2377–83. https://doi.org/10.1093/mnras/stx232.

Rein, H., D. Tamayo, S.-F. Liu, L. Winkler, P. Bartram, A. Silburt, G. Brown, et al. 2025. "Hannorein/Rebound: 4.4.10." Zenodo. https://doi.org/10.5281/zenodo.15784767.

Turk, M. J., B. D. Smith, J. S. Oishi, S. Skory, S. W. Skillman, T. Abel, and M. L. Norman. 2011. "yt: A Multi-code Analysis Toolkit for Astrophysical Simulation Data." *The Astrophysical Journal Supplement Series* 192 (January): 9. https://doi.org/10.1088/0067-0049/192/1/9.

Williams, Thomas, and Colin Kelley. 2013. "Gnuplot 4.6: An Interactive Plotting Program." http://gnuplot.sourceforge.net/.

Williams, Jonathan Carifio, Henrik Norman, and A. David Weigel. 2022. "A Novel JupyterLab User Experience for Interactive Data Visualization." *arXiv e-Prints*, December, arXiv:2212.03907. https://doi.org/10.48550/arXiv.2212.03907.

**Footnotes**

1. A WebAssembly compiler is required. ↩
2. REBOUND also comes in the form of a pre-compiled Python package in which case no compiler is needed. ↩
3. At HD (Retina) resolution. ↩
4. For more details on Cross Origin Opener Policy (COOP) and Cross Origin Embedder Policy (COEP) headers see the emscripten documentation. Whereas we could easily set these policies in our own web server implementation, it is currently not possible to set them on many hosting providers such as ReadTheDocs and GitHub pages. ↩
5. Because Windows APIs for sockets are very different from UNIX APIs, the code for these operating systems is quite different. ↩
6. REBOUND uses simplefont, a low resolution texture atlas. ↩

7. REBOUND handles a failed attempt to connect to the server by retrying with exponentially increasing [delays](). ↵